\documentclass[twocolumn, apj]{ojap}

\usepackage{xspace, color}
\usepackage[breaklinks,colorlinks,citecolor=blue,urlcolor=blue]{hyperref}

\newcommand{\fullname}{LAMOST~J235456.73+335625.9\xspace}
\newcommand{\name}{J2354\xspace}

\newcommand{\um}{\hbox{\ensuremath{\mu \rm m}}\xspace}
\newcommand{\kms}{\hbox{\ensuremath{\rm{km}~\rm s^{-1}}}\xspace}
\newcommand{\Msun}{\hbox{\ensuremath{\rm M_\odot}}\xspace}
\newcommand{\Rsun}{\hbox{\ensuremath{\rm R_\odot}}\xspace}
\newcommand{\AAA}{\hbox{\ensuremath{\rm \AA}}\xspace}

\newcommand{\Ha}{\hbox{H$\alpha$}\xspace}

\newcommand{\kd}{K-dwarf\xspace}
\newcommand{\Kd}{K dwarf\xspace}

\newcommand{\ccsn}{CC~SN\xspace}

\newcommand{\Mstar}{\hbox{\ensuremath{M_\star}}\xspace}
\newcommand{\Teff}{\hbox{\ensuremath{T_{\rm eff}}}\xspace}
\newcommand{\logg}{\hbox{\ensuremath{\log g}}\xspace}
\newcommand{\MH}{\hbox{\ensuremath{ [\rm M/\rm H] }}\xspace}

\newcommand{\FeH}{\hbox{\ensuremath{ [\rm{Fe}/ \rm H] }}\xspace}

\newcommand{\Rstar}{\hbox{\ensuremath{R_\star}}\xspace}
\newcommand{\Mzams}{\hbox{\ensuremath{M_{\rm ZAMS}}}\xspace}

\newcommand{\vsini}{\hbox{\ensuremath{v_{\rm rot}\sin i}}\xspace}

\newcommand{\Mco}{\hbox{\ensuremath{M_{\rm co}}}\xspace}

\newcommand{\fcap}{\hbox{\ensuremath{f_{\rm cap}}}\xspace}
\newcommand{\teff}{\hbox{\ensuremath{T_{\rm eff}}}\xspace} 

\newcommand{\korg}{\texttt{Korg}\xspace} 

\newcommand{\Twd}{\hbox{\ensuremath{T_{\rm WD}}}\xspace}
\newcommand{\Rwd}{\hbox{\ensuremath{R_{\rm WD}}}\xspace}
\defcitealias{zheng2023}{Z23}

\begin{document}

\title{Weighing The Options:\\The Unseen Companion in LAMOST~J2354 is Likely a Massive White Dwarf}

\shorttitle{The Unseen Companion in LAMOST~J2354}
\shortauthors{M. A. Tucker et al.}

\author{
\vspace{-1.3cm}
M. A. Tucker$^{1,2,\star}$}
\author{A. J. Wheeler$^2$}
\author{D. M. Rowan$^{1,2}$}
\author{M. E. Huber$^3$}

\affiliation{$^1$Center for Cosmology and AstroParticle Physics, 191 W Woodruff Ave, Columbus, OH 43210}
\affiliation{$^2$Department of Astronomy, The Ohio State University, 140 W 18th Ave, Columbus, OH 43210}
\affiliation{$^3$Institute for Astronomy, University of Hawai`i, 2680 Woodlawn Drive, Honolulu HI 96822}

\altaffiliation{$^\star$CCAPP Fellow}
\email{Corresponding author: tuckerma95@gmail.com}



\begin{abstract}

\fullname (J2354) is a binary system hosting a $\sim 0.7~\Msun$ K dwarf and a $\sim 1.4~\Msun$ dark companion, supposedly a neutron star, in a 0.48~d orbit. Here we present high- and low-resolution spectroscopy to better constrain the properties of the system. The low-resolution spectrum confirms that the luminous star is a slightly metal-poor K dwarf and strengthens the limits on any optical flux from the dimmer companion. We use the high-resolution spectra to measure atmospheric parameters ($\teff, \logg, \FeH, \vsini$) and abundances for 8 elements for the \Kd. We refine the mass of the compact object to $M_{\rm co} \sim 1.3$~\Msun with a minimum mass of $M_{\rm co, min} = 1.23\pm0.04$~\Msun. The expected overabundance of intermediate-mass elements from the incident supernova ejecta is not detected in the \Kd atmosphere. This contrasts with known binaries hosting neutron stars where almost all companions show evidence for polluting material. Moving the neutron-star progenitor further from the K-dwarf at the time of explosion to minimize atmospheric pollution requires a finely-tuned kick to produce the current orbital separation of $\sim 3.3~\Rsun$. Instead, we find that a massive white dwarf with a cooling age of $\gtrsim 3~$Gyr satisfies all observational constraints. The system likely experienced two common-envelope phases leading to its current state because the white dwarf progenitor was massive enough to ignite He-shell burning. The system will become a cataclysmic variable in the distant future when the K-dwarf evolves off of the main sequence. These short-period high-$q$ binaries represent an intriguing formation pathway for compact double white dwarf binaries and thermonuclear supernovae. An ultraviolet spectrum is the most promising avenue for directly detecting the white dwarf companion.
\end{abstract}

\keywords{Close binary stars (254), White dwarf stars (1799), Common envelope evolution (2154), Low mass stars (2050), Stellar abundances (1577)}


\section{Introduction} \label{sec:intro}

Massive white dwarfs (WDs) are the remnants of $6-8~\Msun$ zero-age main sequence (ZAMS) stars. These stars are rare, confirmed by the rarity of massive WDs \citep[e.g., ][]{kilic2021}, but represent a key contributor of nucleosynthetic material to the interstellar medium (ISM). A single $\sim 7~\Msun$ star contributes the same amount of material to the ISM as twenty 1-\Msun stars \citep{cummings2018}. The core composition of massive WDs traces which elements were fused during stellar evolution. Stars with $\Mzams \sim 8-12~\Msun$ can ignite carbon but not neon in the core, producing O+Ne+Mg WDs instead of the C/O WDs that originate from stars with $\Mzams\sim 0.5-8~\Msun$ \citep[e.g., ][]{camisassa2019}. Characterizing these stellar remnants constrains relativistic effects in degenerate plasmas \citep{althaus2023}, traces the star-formation of the Milky Way \citep{fantin2019}, and reveals the chemical composition of post-MS planetary systems \citep{jenkins2024}.

Yet these intermediate-mass stars that become massive WDs prefer companionship, with more than half having a companion close enough for mass-transfer during their evolution \citep{moe2017}. This will affect the long-term evolution of both stars, complicating reliable comparisons to WD cooling models. Close binaries will almost always experience a common-envelope (CE) phase where the lower-mass star becomes embedded in the extended, tenuous envelope of the more massive star as it evolves through the giant branch (see \citealp{ropke2023} for a recent review). CE interaction transfers orbital energy to the envelope, producing tighter binaries that are the precursors for stellar mergers, supernova progenitors, and X-ray binaries. Despite the ubiquity of mass transfer in binary stellar evolution and its wide-ranging influence across astrophysical disciplines, we have only a crude understanding of the physical processes determining the final outcome(s) \citep[e.g., ][]{ivanova2011, postnov2014, marchant2021, belloni2023}.

Discovering and characterizing post-CE binaries remains the most promising avenue for empirically constraining the underlying physics \citep[e.g., ][]{zorotovic2010, zorotovic2022,scherbak2023, belloni2024, yamaguchi2024b}. Yet discovering these systems is complicated by the small sizes and inherently low luminosities of WDs, especially since even low-mass K and M dwarfs can outshine the WD at optical and infrared (IR) wavelengths \citep[e.g., ][]{RM2021}. Identifying WDs in binaries was historically biased towards accreting systems, such as cataclysmic variables (CVs), because accretion produces luminous emission spanning the electromagnetic spectrum \citep[e.g., ][]{ritter2003, barlow2006, drake2014, schwope2024}. This is shifting in the era of astrometry and parallaxes from the \emph{Gaia} mission which has revolutionized the discovery of WDs in wider binaries \citep[e.g., ][]{shahaf2024,yamaguchi2024}. 

Yet there exists a subset of quiescent WD+main sequence (MS) binaries with small separations ($a\lesssim 100~\Rsun$). They are not close enough to drive accretion but too close for \emph{Gaia} to detect photocenter variability from the orbital motion. Lower-mass WDs have larger radii and thus higher luminosities, increasing their chances of detection \citep[e.g., ][]{kosakowski2020}. There is also a bias towards young WDs with red companions (i.e., M dwarfs) as each star contributes similar luminosities at different wavelengths (e.g., \citealp{RM2012}). More difficult to find are the smaller and more massive WDs, which also cool faster than their lower-mass counterparts \citep{bedard2020, camisassa2022}. Careful analyses of astrometric, spectroscopic, and photometric observations are typically required to confidently identify these systems \citep[e.g., ][]{rowan2024}. 

Further complicating the search for massive WDs in quiescent binaries is the overlap between massive WDs ($< 1.4~\Msun$; \citealp{takahashi2013}) and low-mass neutron stars (NSs, $\gtrsim 1.2~\Msun$; \citealp{ozel2012, suwa2018}). Young NSs can typically be identified in radio or X-ray observations \citep[e.g., ][]{atnf_cat, integral_cat} and young WDs are often hot ($\teff \gtrsim 10^4$~K) making them bright in the ultraviolet \citep[UV; e.g., ][]{gray2011, parsons2016, garbutt2024}. Older systems with slowly-spinning NSs or cool WDs have limited options for conclusively determining the nature of a $\sim 1.2-1.4~\Msun$ unseen companion to a more luminous MS star.

Finding stars with large radial velocity (RV) variations remains a promising avenue for detecting quiescent compact objects \citep[e.g., ][]{jayasinghe2023, liu2024, rowan2024b}. \citeauthor{zheng2023} (\citeyear{zheng2023}, hereafter \citetalias{zheng2023}) recently reported the discovery of \fullname (\name), a nearby ($d=127.7\pm0.3$~pc) K dwarf with a $\sim 1.4-1.6~\Msun$ unseen companion in a $\sim 0.48$~d orbit. They favor a NS companion but note that a massive WD cannot be fully excluded. After its discovery, we obtained follow-up spectroscopy of \name to refine the inferred masses and search for polluting ejecta in the \Kd atmosphere. Here we find that the unseen companion is more likely to be a massive WD instead of a NS. The data reduction and calibration are described in \S\ref{sec:data}. We measure atmospheric parameters (\Teff, \logg, \FeH, \vsini) and the abundances of eight elements for the \Kd in \S\ref{sec:kd}. We show in \S\ref{sec:discuss} that a massive WD is more plausible than a low-mass NS based on available observations. Finally, \S\ref{sec:summary} summarizes our results. 

\section{Spectroscopic Observations}\label{sec:data}

\begin{figure*}
    \centering
    \includegraphics[width=\linewidth]{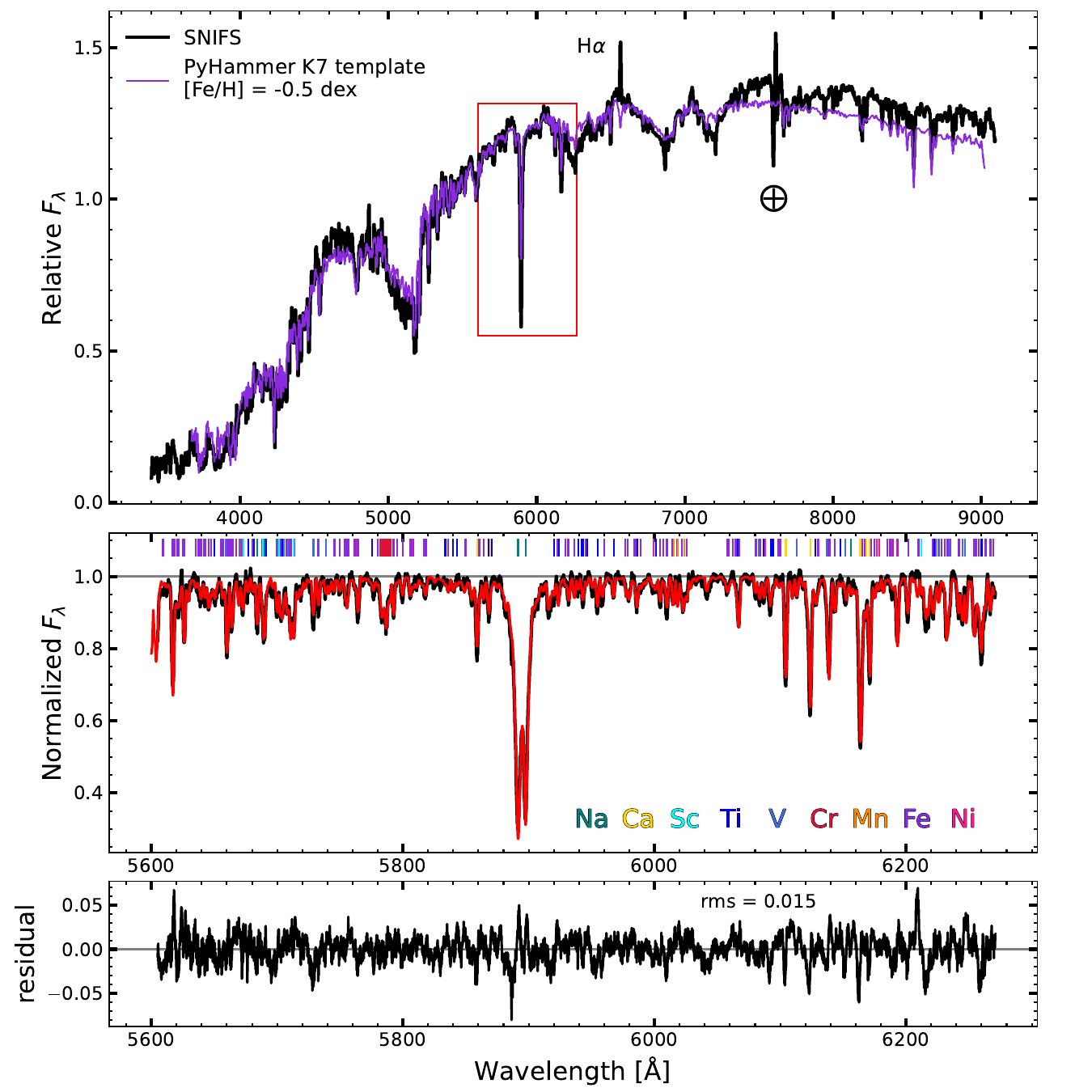}
    \caption{Spectroscopic observations of \name. The top panel shows the $R\sim 1200$ SNIFS spectrum compared to the best-match [M/H]$=-0.5$~dex PyHammer \citep{pyham1, pyham2} K7V template (purple), convolved and re-binned to the SNIFS resolution. The red rectangle shows the wavelength coverage of the $R\approx 130,000$ PEPSI spectrum shown in the middle panel. The best-fit atmospheric model is overlaid in red and the residuals are shown in the bottom panel. The PEPSI spectrum and residuals have been re-binned to $0.1~\rm\AA$/pixel for visual clarity. Colored ticks along the top of the middle panel denote features used to measure abundances.}
    \label{fig:all-spex}
\end{figure*}

We obtained low-resolution and high-resolution follow-up spectroscopic observations of \name after its discovery. The low-resolution spectrum was obtained by the SuperNova Integral Field Spectrograph \citep[SNIFS; ][]{lantz2004} on the UH2.2m telescope through the Spectroscopic Classification of Astronomical Transients \citep[SCAT; ][]{tucker2022} survey. The spectrum covers $\approx 3400-9000$~\AAA at a resolution of $R \approx 1200$.  

Three high-resolution spectra were obtained with the Potsdam Echelle Polarimetric and Spectroscopic Instrument \citep[PEPSI; ][]{pepsi1, pepsi2} on the Large Binocular Telescope (LBT) under good conditions. We used the 200~\um fiber yielding a resolution of $R\approx 130,000$ ($\Delta v \approx 2.3~\kms$). The CD2 and CD4 dispersers were used to cover $4300-4800~\AAA$ and $5500-6300~\AAA$ simultaneously. Table~\ref{tab:speclog} provides exposure information for the PEPSI observations. 

We measure and report radial velocities (RVs) for both channels in Table~\ref{tab:speclog} but only the red channel is used for determining stellar parameters due to the strong line-blanketing at blue ($\lesssim 5500~\AAA$) wavelengths. This prevents reliably measuring the stellar continuum, and attempts to fit both channels simultaneously were unsuccessful. Using only the red channel also allows us to ignore potential contamination at blue wavelengths from a WD or low-luminosity accretion disk. 

\begin{table*}[]
    \centering
    \caption{Information for the three 40-minute PEPSI exposures. RVs are given in \kms. Phases are computed using the \citetalias{zheng2023} ephemeris. The Barycentric Dynamical Time Julian Date (JD-TBD) corresponds to the middle of each exposure. Radial velocities are measured by cross-correlating the observed spectrum and a synthetic template with the derived spectroscopic parameters reported in Table \ref{tab:korg_bulk} in {\tt iSpec} \citep{blancocuaresma14}. \\
    $^a$Median SNR per exposure for the blue/red channel.}
    \label{tab:speclog}

    \begin{tabular}{lcccccccc}
    \hline\hline
    Exposure Start (UTC) & JD$-$TDB & Phase & medSNR$^a$ & RV$_{\rm blue}$ & RV$_{\rm red}$ & RV$_{\rm Z23}$ \\ 
    \hline
    2022-10-14 02:19:38 & 2459866.616501445 & $0.33-0.39$ & $27/86$ & $210.4\pm6.6$ & $204.5\pm5.0$ & $212.9$ \\
    2022-10-14 03:00:25 & 2459866.644826776 & $0.27-0.33$ & $32/98$ & $247.4\pm6.0$ & $243.5\pm3.4$ & $250.6$ \\
    2022-10-14 03:41:13 & 2459866.673155239 & $0.21-0.27$ & $34/101$ & $255.6\pm5.7$ & $250.7\pm3.6$ & 259.8 \\
    \hline\hline
    \end{tabular}
\end{table*}

\section{Properties of the K dwarf}\label{sec:kd}

\begin{table}[]
    \centering
    \caption{The adopted stellar parameters and uncertainties based on the high-resolution PEPSI observations.  See text for details.}
    \label{tab:korg_bulk}

    \begin{tabular}{lccccc}
    \hline\hline
    Parameter & Value & $\sigma_\mathrm{sys}$ & $\sigma_\mathrm{stat}$ & $\sigma_\mathrm{tot}$ & Unit \\\hline
    \Teff &  4327 & 100 & 11 & 101 & K \\
    \logg &  4.66 & 0.20 & 0.082 & 0.22 & dex \\ 
    \MH &  -0.48 & 0.10 & 0.026 & 0.10 & dex \\ 
    \vsini &  70.3 & 2.7 & 2.8 & 3.9 & \kms \\ 
    \hline\hline
    \end{tabular}
\end{table}

\subsection{Bulk Properties}

Fig.~\ref{fig:all-spex} shows the spectra of \name. We use PyHammer \citep{pyham1, pyham2} to compare the low-resolution spectrum against a library of templates. The best-matching template is a metal-poor ($\FeH=-0.5$~dex) K7 dwarf which we use to initialize the stellar parameters when fitting the high-resolution spectrum. We adopt the photo-geometric distance of $d = 127.3\pm0.3$~pc from \citet{bailerjones2023} and interstellar reddening $A_V = 0.00^{+0.03}_{-0.00}$~mag based on the dust maps of \citet{green2019} to remain consistent with \citetalias{zheng2023}.\footnote{The more recent 3D dust maps of \citet{mwdust} predict a non-negligible extinction of $A_V = 0.12^{+0.05}_{-0.04}$~mag. This does not affect the interpretation of \name besides a slight increase to the UV excess.}

We analyze the high-resolution PEPSI spectrum using the \texttt{fit\_spectrum} function in the \korg{} \citep{wheeler2023, wheeler2024} spectral synthesis package.
First, we simultaneously fit \teff, \logg, $\MH$, and $\vsini$ to the normalized PEPSI spectrum. The empirical continuum normalization is imperfect, so we include minor continuum corrections after the first fit and re-fit with the adjusted continuum.
The linelist was obtained from the ``extract stellar'' mode of the Vienna Atomic Line Database (VALD; \citealp{piskunov1995, ryabchikova2015}).\footnote{The ``extract stellar'' mode is designed to provide all known spectral features in the specified $\lambda$ range for a given set of stellar parameters. Note that VALD includes both atomic and molecular lines.} The linelist references can be found in Appendix \ref{app:obs}.

The high signal-to-noise ($\sim 100$) and spectral resolution ($R\sim 130,000$) of the PEPSI spectra make the systematic uncertainties dominant over statistical ones. These systematics include physical uncertainties in stellar atmosphere models, computational shortcuts in radiative transfer modeling (i.e., LTE), and instrumental/observational effects such as orbital smearing. For \teff, \logg, and \MH we estimate systematic uncertainties of $100$~K, 0.2~dex, and $0.1$~dex based on \citet{jofre2019} and \citet{hegedus2023}. We propagate these uncertainties to rotation by refitting for \vsini{} with each parameter perturbed in each direction by its assumed systematic uncertainty and adding the perturbations to \vsini{} in quadrature. The uncertainty in \MH dominates the systematic uncertainties in \vsini.

Individual fits to the three 40-minute exposures provide stellar parameters very close to those from the stacked-spectrum fit, but with dispersions slightly larger than the statistical errors on the parameters. We adopt the sample standard deviation among the visits as the measurement error for each parameter. Measurement of \vsini is complicated by potential orbital smearing during the 40-minute exposures. The orbital ephemeris specified by \citetalias{zheng2023} suggests the spectra span orbital phases of $\phi \approx 0.2-0.4$\footnote{$\phi\equiv0$ corresponds to the visible star in superior conjunction.}  where radial acceleration is minimal, and the period uncertainty of $10^{-5}$~d corresponds to a phase uncertainty of $\delta \phi\approx 0.1$. 
In order to better account for orbital smearing, we adopt the SNR-weighted average of the \vsini{} values from each fit, which is 2.9 \kms lower than that from the stacked spectrum.

The final atmospheric parameters and adopted uncertainties are provided in Table~\ref{tab:korg_bulk}. Our values generally agree with those published by \citetalias{zheng2023} within uncertainties.\footnote{
We inflate their uncertainties by the average model `spread' reported in Table~6 of \citet{vines2022}: $\delta T_{\rm eff} = 3\%$, $\delta \logg = 0.1$~dex, $\delta \Rstar = 8\%$, and $\delta \FeH = 0.2$~dex.
} Overall, the luminous star is a rapidly-rotating metal-poor K-dwarf. We estimate $\Mstar = 0.61\pm 0.04~\Msun$ and $\Rstar=0.65\pm0.03$ by fitting the derived \teff and \logg to MIST isochrones \citep{mist1, mist2}. \citetalias{zheng2023} find a similar \Rstar but a higher $\Mstar = 0.73\pm0.05~\Msun$ using \logg and \Rstar from their SED fit. The infrared mass-luminosity-metallicity relation derived by \citet{mann2019} predicts $\Mstar = 0.63\pm0.02~\Msun$. Given the unknown nature of the unseen companion, potential for a low-luminosity accretion disk, and likely contamination from the \Kd chromosphere, we adopt $\Mstar = 0.65\pm0.05$~\Msun and $\Rstar = 0.65\pm0.05~\Rsun$ to span all estimates.

\begin{figure}
    \centering
    \includegraphics[width=\linewidth]{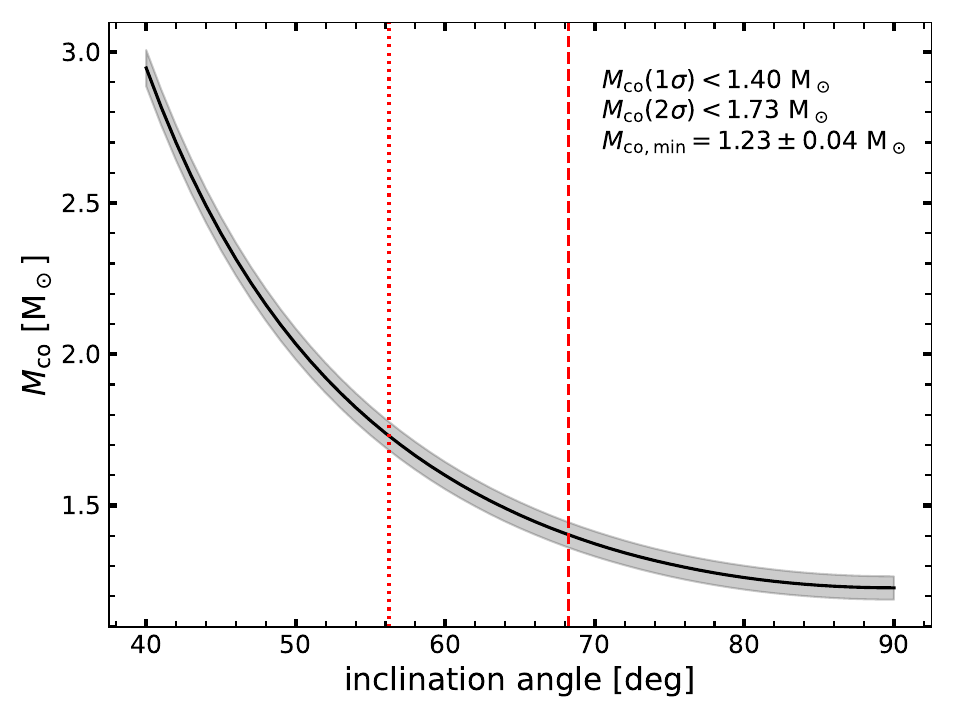}
    \caption{We derive $\sin i = 1.03\pm0.10$ from the measured \vsini and \Rstar. $i>90$~deg is unphysical so we report $1\sigma$ and $2\sigma$ bounds on the mass of the compact object, \Mco, in conjunction with the lower limit required by $i\equiv90$~deg.
    }
    \label{fig:incl}
\end{figure}

The timescale for circularizing and synchronizing the orbit is only $\approx 10^5$~yr \citep{zahn1977} and the low eccentricity ($e = 0.002\pm0.002$) reported by \citetalias{zheng2023} agrees with a tidally-locked binary. Tidal synchronization allows a direct mass estimate for the unseen companion because $v_{\rm rot} = 2\pi \Rstar/P$ and $v_{\rm rot}\sin i = 70.3\pm3.9~\kms$ (Table~\ref{tab:korg_bulk}). We solve for the inclination angle using our \Rstar estimate and the period derived by \citetalias{zheng2023} (${P = 0.47992 \pm 0.00001}$~d), finding $\sin i = 1.03\pm0.10$. We propagate the uncertainties in \Rstar and $P$ to $\sin i$ using a Markov Chain Monte Carlo routine. ${\sin i > 1}$ implies $i>90$~deg which is not physically reasonable, either due to an underestimated \Rstar or an overestimated \vsini. \citetalias{zheng2023} find no eclipses in the TESS light curve of \name, requiring $i< 90^\circ$, but the small radii of massive WDs ($R_{\rm WD}\lesssim 0.01~\Rstar$) means this constraint is very weak (cf. Fig.~11 in \citealp{rowan2024}). We report $1\sigma$ and $2\sigma$ limits on the mass of the compact object, \Mco, in Fig.~\ref{fig:incl} alongside the minimum \Mco assuming $i\equiv90^\circ$. We adopt $\Mco = 1.30_{-0.05}^{+0.10}~\Msun$ for the analysis in \S\ref{sec:discuss} which is marginally below the lower estimate quoted by \citetalias{zheng2023} but consistent within uncertainties. The corresponding orbital separation is $a = 3.3\pm0.1~\Rsun$.

\subsection{Individual Abundances}\label{subsec:kd.abun}
\begin{table}[]
    \begin{center}
    \caption{Adopted abundance and uncertainties for each element measured from the high-resolution PEPSI spectra. A systematic uncertainty of 0.1 dex is assumed for each (see \S\ref{subsec:kd.abun}). Abundances are given relative to Solar \citep{asplund2021}.}
    \label{tab:korg_abunds}

    \begin{tabular}{lccc}
    \hline\hline
    Element & [$X$/H] (dex) & $\sigma_\mathrm{stat}$ & $\sigma_\mathrm{tot}$ \\\hline
    Na & -0.28 & 0.20 & 0.23 \\
    Ca & -0.37 & 0.02 & 0.10 \\
    Sc & -0.60 & 0.19 & 0.22 \\
    Ti & -0.44 & 0.06 & 0.12 \\
    V & -0.38 & 0.08 & 0.13 \\
    Cr & -0.39 & 0.05 & 0.11 \\
    Mn & -0.66 & 0.04 & 0.11 \\
    Fe & -0.43 & 0.02 & 0.10 \\
    Ni & -0.23 & 0.14 & 0.17 \\
    \hline\hline
    \end{tabular}
    \end{center}
\end{table}

We use \korg{}'s \texttt{fit\_spectrum} functionality to measure elemental abundances for \name.
To identify features strong enough to measure elemental abundances, we used \korg{}'s \texttt{prune\_linelist} function (with the threshold ratio in line-center absorption to continuum absorption set to 1.0).
We then visually inspected the observed and synthetic spectra to narrow down the list of measurable elements.
We first fit the whole spectrum with the initial stellar parameters held fixed to initialize a fiducial value for all elements.
Then, each feature was fit within a 10~\AA{} window while holding the other abundances fixed.
The average abundance and error on the mean for each element are listed in Table~\ref{tab:korg_abunds}. 
Features where excluded from this calculation if they had statistical error greater than one, if they were nondetections (defined as a best-fit $[X/\mathrm{H}] < -3$) or (for elements with more than 20 features) if they overlap with the strong sodium D absorption lines.
Elements with fewer than three remaining features (Mg and Ba) were excluded.
For all elements, the line-to-line scatter is larger than the statistical uncertainty in the per-line abundance.  
We adopt the sample standard deviation of the line-to-line estimates as the measurement error, which we add in quadrature to a systematic uncertainty of 0.1 dex for each element \citep[e.g., ][]{jofre2019}.
Table \ref{tab:korg_abunds} lists the abundances and uncertainties for each element, and Appendix \ref{app:obs} contains further details.

Fig.~\ref{fig:abund_compare} compares the abundances of \name to stars within 100~pc of it based upon the photogeometric distances of \citet{bailerjones2023} that have abundances from the GALactic Archaeology with HERMES (GALAH; \citealp{galah}) survey.\footnote{Using abundances from the APOGEE experiment \citep{abdurrouf2022} instead of GALAH does not meaningfully change our conclusions.} We also show abundances for systems with confirmed or high-confidence compact object companions to search for unique elemental signatures or trends including WDs \citep{kong2018a, kong2018b}, NSs \citep{GH2005,hinkle2006,hinkle2019,hinkle2020,SA2015,shanhbaz2022,gaiaNS1} and BHs \citep{israelian1999,GH2004,GH2006,GH2008,GH2011,sadakane2006,gaiaBH1, gaiaBH2, gaiaBH2-2, gaiaBH3}. Evolved stars ($\logg < 3.0$~dex) are highlighted because they may experience different mixing and dilution properties than dwarfs. Interestingly, systems that experienced a core-collapse supernova often show evidence for pollution traced by enhanced $\alpha$ or intermediate-mass elements (IMEs). We note that Fig.~\ref{fig:abund_compare} obscures the dependence on orbital separation.  

\section{Implications for the Unseen Companion}\label{sec:discuss}

\subsection{Neutron Star Scenarios}\label{subsec:discuss.NS}

\begin{figure*}
    \centering
    \includegraphics[width=\linewidth]{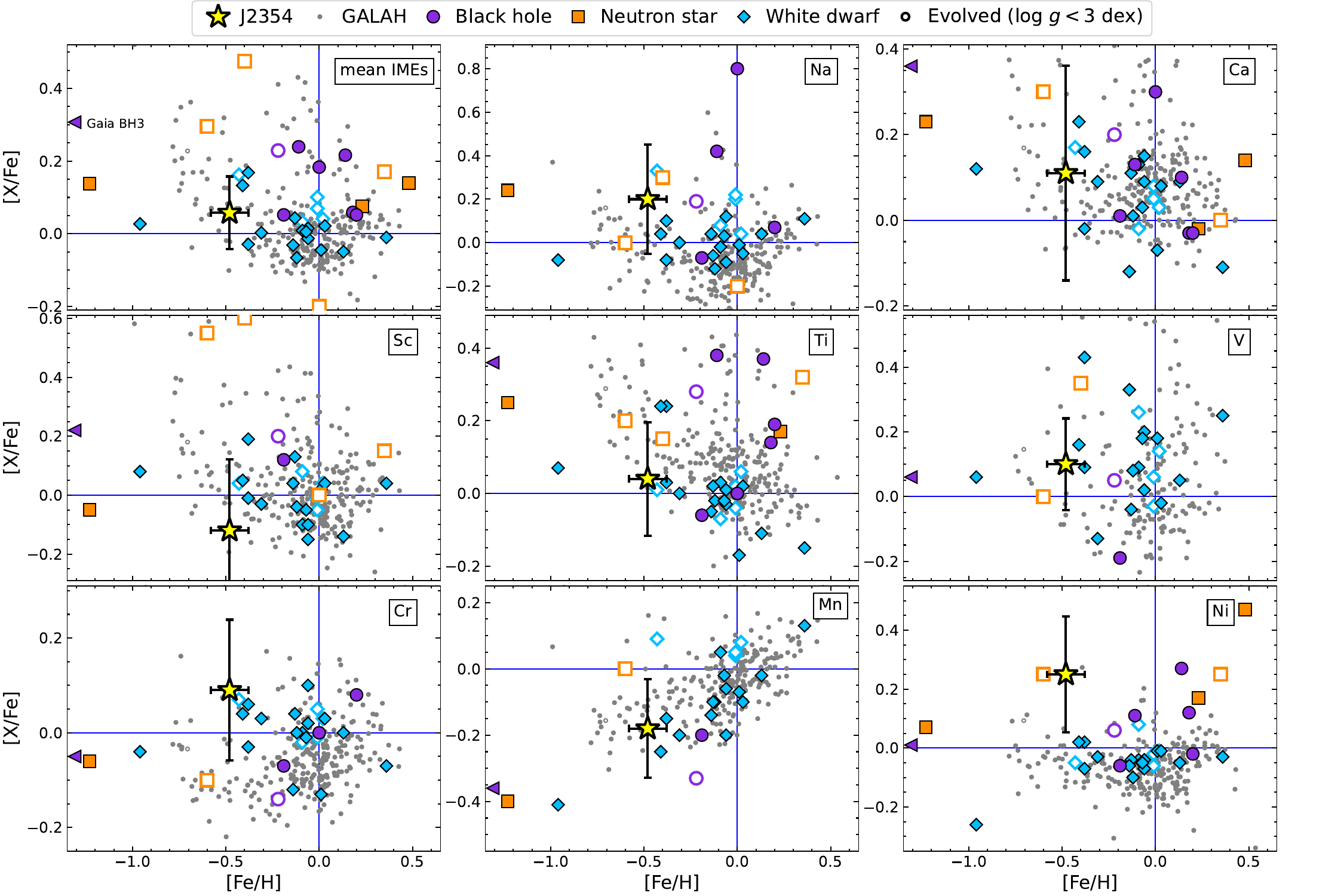}
    \caption{The elemental abundances derived for \name and the mean IMEs computed from Na, Ca, Ti, and Sc. Stars within 100~pc of \name with GALAH DR3 abundances are shown as small gray points. We also show known binary systems hosting black holes (purple), neutron stars (orange) and white dwarfs (blue). Evolved companions ($\log g < 3$~dex) are shown with open symbols. The purple triangle is the low-metallicity system Gaia~BH3 ($\FeH\approx -2.5$~dex) which we have shifted horizontally by +1.25~dex for visual clarity. Abundance uncertainties for the comparison sources are typically similar to those we derive for \name. Thin blue lines mark Solar abundance ratios. References are provided at the end of \S\ref{subsec:kd.abun}.}
    \label{fig:abund_compare}
\end{figure*}

For a spherical explosion, the amount of polluting material deposited onto the companion's surface can be estimated assuming that it is proportional to the solid angle subtended by the companion, 

\begin{equation}\label{eq:dmcap}
    \Delta m_{\rm cap} = f_{\rm cap} \times \Delta M \times (\pi \Rstar / 2\pi a_0)^2
\end{equation}

\noindent where $\Delta m_{\rm cap}$ is the mass of polluting material accreted by the companion, $\Delta M$ is the supernova (SN) ejecta mass, $R_\star$ is the companion radius, $a_0$ is the pre-explosion orbital separation, and $f_{\rm cap}$ is a scaling factor representing the fraction of incident ejecta that stays bound to the \Kd such that $f_{\rm cap} = 0-1$. Simulations show that these simple estimates are generally reasonable - the lower-mass companion will have some mass ablated from the surface which reduces \fcap for the outer H-rich ejecta but increases \fcap for the slower-moving inner ejecta \citep{liu2015}.

\name does not show obvious evidence for pollution in Fig.~\ref{fig:abund_compare} which disfavors a nearby \ccsn. In most systems that host a NS or BH the companion shows evidence for atmospheric pollution traced by above-average abundances of IMEs. Variations in ejecta pollution are likely driven by different orbital distances, especially given that the current orbits are not the pre-SN orbits \citep[e.g., ][]{brandt1995} and the systems in Fig.~\ref{fig:abund_compare} have a variety of orbital separations. The explosion energy and ejecta mass of the SN will introduce smaller variations in companion pollution, as will mass- and metallicity-dependent nucleosynthesis \citep[e.g., ][]{heger2010, limongi2018}.

The current orbit is not equal to the pre-SN orbit due to mass ejection by the explosion and the potential for NS kicks \citep[e.g., ][]{sweeney2022}. Moreover, the expected effects of a SN on a $\sim 0.65$~\Msun star orbiting just $\sim 3.3~\Rsun$ away are severe. The impacting ejecta will ablate $\gtrsim 10\%$ of the \Kd mass, impart a $\sim 100~\kms$ kick to the \Kd, and deposit $\sim10^{-3}$~\Msun of material onto the surface \citep{liu2015}, which is in tension with our abundance analysis. Taking the nucleosynthetic yield of a $15~\Msun$ progenitor from \citet{sukhbold2016}, the accreted mass estimates as a function of orbital separation from \citet{liu2015}, and $f_{\rm cap} = 0.1$, the ejecta would have contributed $\sim 10\%$ of the total iron-group elements (Mn, Fe, Ni) by mass and $\sim 25\%$ of the IMEs (Na, Ca, Sc, Ti). This would require an IME-poor \Kd prior to the SN, at-odds with known trends of increasing $\alpha$ elements/IMEs with decreasing \FeH \citep[e.g., ][]{hayden2015}. 

These tensions can be mitigated by increasing the pre-SN orbital separation to decrease the pollution and allow a more normal abundance pattern in the \Kd (cf. Fig.~\ref{fig:abund_compare}) prior to the explosion. However, most binaries will expand their orbit after one component explodes due to mass loss from the system. A natal kick of order or higher than the pre-SN orbital velocity is needed to produce a tighter binary \citep{brandt1995, kalogera1996}. Yet less than a third of systems survive such a randomly-oriented kick and only $\sim 10\%$ will produce tighter binaries. Thus, there must be a balance between increasing the pre-SN separation to reduce atmospheric pollution of the \Kd and decreasing the orbital separation to improve the chance of producing the correct post-SN orbit.\footnote{We ignore the kick imparted onto the companion by the SN \citep{liu2015} as it only decreases the viability of a surviving binary.}

These are indirect arguments against the unseen companion being a quiescent NS, especially when considering theoretical uncertainties in massive-star evolution, nucleosynthesis, explosion sphericity, and so on. Yet the circumstantial evidence is accumulating given the delicate balance between reducing atmospheric pollution and requiring a tight post-SN orbit. Moreover, a quiescent NS+MS system at $\approx 127$~pc is a factor of $\sim 2$ closer than any quiescent NS+MS system reported by \citet[][$d_{\rm min} \sim 250$~pc]{elbadry2024}. This corresponds to $1/8$th of the search volume, or an increase in the predicted number of quiescent NSs in binaries by almost an order of magnitude. Given the inconclusive evidence for a NS as the dim companion, we reevaluate this assumption in the next section.

\subsection{White Dwarf Scenarios}

\begin{figure*}
    \centering
    \includegraphics[width=\linewidth]{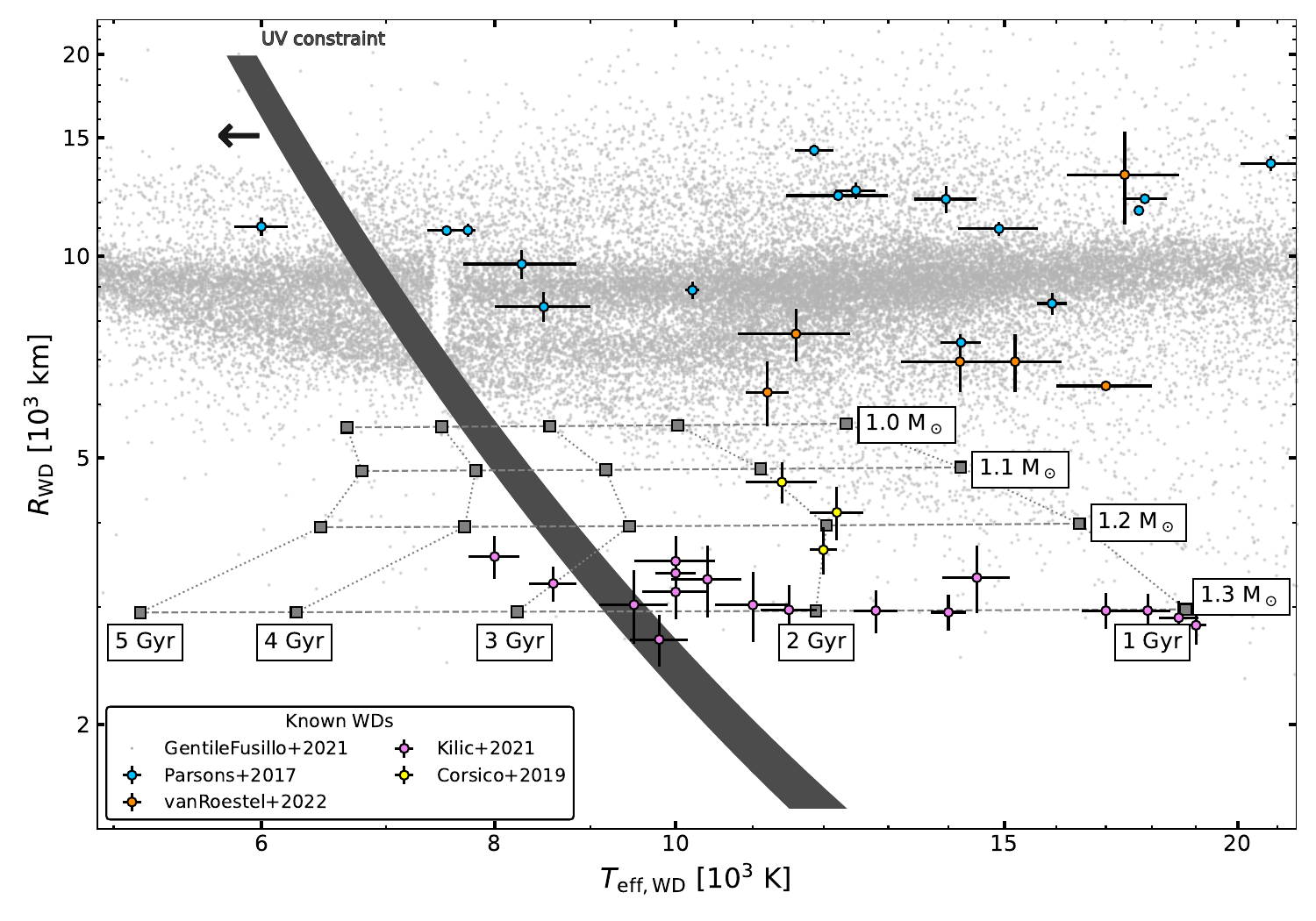}
    \caption{\Teff versus radius for a putative WD companion. The dark-gray region + arrow shows the constraint from the averaged GALEX NUV and \emph{Swift} UVM2 photometry. The combined UV flux from the WD + \Kd chromosphere must equal this constraint. For comparison, we show known WDs \citep[bold points, ][]{parsons2017, corsico2019, kilic2021, vanRoestel2022} and WD candidates from Gaia EDR3 \citep[small gray points, ][]{GF2021}. Overlaid are conservative cooling sequences for massive C/O WDs (bold gray squares) computed by \citet{bedard2020} showing a $\sim 1.3~\Msun$ WD with a cooling age of $\gtrsim 3$~Gyr is allowed. Models adopting O+Ne+Mg cores or thinner H envelopes cool faster, and models including the effects of phase separations or general relativity have even smaller radii \citep{camisassa2019, camisassa2022, althaus2022, althaus2023}.}
    \label{fig:Twd-Rwd}
\end{figure*}

\citetalias{zheng2023} show that the \emph{Swift} UV colors are incompatible with the blackbody-like emission expected for a WD. Instead, they attribute the UV excess to chromospheric activity from the active \kd, supported by \Ha emission in the LAMOST spectra (and seen in the SNIFS spectrum). Yet they do not explore the possibility that the UV excess can be explained by a combination of chromospheric activity and WD emission. 

Fig.~\ref{fig:Twd-Rwd} shows the allowed parameter space given the GALEX and \emph{Swift} near-UV detection ($m_{\rm NUV} = 20.15\pm0.11$~mag) for a blackbody with temperature \Twd and radius \Rwd. Similar constraints can be obtained from the non-detection a blue continuum in the SNIFS spectrum (cf. Fig.~\ref{fig:all-spex}). We use the \textsc{Phoenix} grid of model atmospheres \citep{phoenix} to estimate the small contribution of the K dwarf to the UV flux and find where the combined UV fluxes equal the observed flux, shown in Fig.~\ref{fig:Twd-Rwd} as the dark gray region. This is an upper limit on the WD UV flux because there is likely some, and potentially significant, UV flux from the \Kd chromosphere. Overlaid in Fig.~\ref{fig:Twd-Rwd} are C/O WD evolutionary tracks from \citet{bedard2020}\footnote{\url{https://www.astro.umontreal.ca/~bergeron/CoolingModels/}} showing the increased cooling rate with increasing mass. We show the most conservative cooling tracks, as adopting thinner H layers \citep{bedard2020} or O-Ne-Mg cores \citep{althaus2022} produces faster cooling or smaller radii, respectively. Including relativistic effects can further shrink the radii of massive WDs by $10-20\%$ \citep{althaus2022, althaus2023}.\footnote{\url{http://evolgroup.fcaglp.unlp.edu.ar/TRACKS/UMall.html}}

Thus, a massive WD can certainly be hidden by the \Kd in \name. As noted by \citetalias{zheng2023}, \name would be one of the most massive WDs in a nearby close binary. \name is strikingly similar to the systems identified by \citet{rowan2024} with $\sim 1~\Msun$ WDs orbiting chromospherically-active (spotted) K dwarfs in $\sim 0.5$-day orbits. \name is also similar to LAMOST~J1123 \citep{yi2022} which is a claimed quiescent NS + M dwarf binary in a $0.28$-d orbit. While K/M dwarfs are common secondaries in canonical post-CE binaries ($\Mstar \lesssim 0.8~\Msun$, \citealp{RM2012, parsons2016}), the cool and massive WDs in \name and these similar systems set them apart. 

This system also represents an intriguing mass-loss problem. The progenitor ZAMS mass of a $\sim1.3~\Msun$ WD is $\Mzams \sim 6-8~\Msun$ \citep[e.g., ][]{elbadry2018, cunningham2024}. This system would have an initial mass ratio $q = M_{\rm WD\;progenitor} / M_{\rm K\;dwarf} \sim 10$ and must have lost $6-1.3\approx 4.7~\Msun$ of material during post-MS evolution. Yet the Eddington accretion limit of the \Kd is $\sim 10^{-8}~\Msun/\rm{yr}$ so even for a 10-million-year RG phase, the \Kd accretes $\lesssim 0.01~\Msun$. Simulations generally predict limited accretion during CE evolution but it may depend on the specifics of each system \citep[e.g., ][]{macleod2015, chamandy2018}. Higher progenitor masses increases the envelope mass and shortens the RG phase, whereas allowing the \Kd to accrete significant amounts of material further increases the original mass ratio.

The WD is massive enough for the system to likely experience two CE phases. The envelope is stripped from the more massive companion during the first CE phase, but stars above $\Mzams\sim 2.5~\Msun$ will also undergo He-shell burning. This causes the envelope to expand again to $\sim 100~\Rsun$ \citep{woosley2019, zhang2024} producing a 2nd CE phase (i.e., `Case BB' mass transfer). This system will almost certainly experience another phase of mass-transfer in the (distant) future when the \Kd evolves off of the MS resulting in stable mass-transfer and the formation of a CV. 

\name and similar short-period high-$q$ binaries \citep[e.g., ][]{rowan2024} represent a unique pathway for thermonuclear (Type Ia) supernovae in old stellar populations. The massive WD could accrete matter rapidly enough to ignite C in the core (single-degenerate scenario; \citealp{whelan1973, nomoto1982}), inspiral and merge with the RG or AGB core then explode (core-degenerate scenario; \citealp{ilkov2013, wang2017}), or inspiral but not merge (double-degenerate scenario) to later experience a double detonation \citep[e.g., ][]{livne1990, townsley2019} or merge as a double-WD binary \citep[e.g., ][]{pakmor2012, pakmor2013}. The outcome will depend sensitively on the future orbital evolution during mass-transfer as a CV \citep[e.g., ][]{neunteufel2016}.

Finally, we note that there is currently no direct evidence for the companion being a WD, similar to the arguments against a NS outlined in \S\ref{subsec:discuss.NS}. A spectrum of the UV excess remains the most promising avenue for direct detection \citep[e.g., ][]{hernandez2022}. Chromospheric activity in low-mass stars produces line-dominated UV emission \citep{france2016} which contrasts with the smooth blackbody-like spectrum expected for a cool WD \citep[e.g., ][]{caron2023, wall2023}. \emph{HST} is the only currently-available facility for such a task. Another interesting prospect for future observations is high-resolution spectroscopy covering lighter species such as CNO, especially isotopologues in the near-IR \citep[e.g., ][]{galan2016, galan2017}.

\section{Summary}\label{sec:summary}

We presented follow-up spectroscopy of the enigmatic binary \name to better understand the \Kd and its unseen, massive companion. Overall, the spectra reveal a relatively uninspiring field \Kd ($\Mstar = 0.65\pm0.05~\Msun$, $\Rstar = 0.65\pm0.05~\Rsun$, $\FeH = -0.48\pm0.10$~dex) with no peculiar abundances. The improved \vsini provides new constraints on the mass of the compact object of $\Mco \sim 1.3~\Msun$ with a minimum of $M_{\rm co,min} = 1.23\pm0.04~\Msun$. It is difficult to reconcile the normal abundance profile of the \Kd and the current close-in orbit ($a = 3.3\pm0.1~\Rsun$) with a NS born from a \ccsn. Instead, we find a massive WD the more plausible scenario. 

Such a system represents a unique view into close binary evolution at high mass ratio ($q\approx 10$) as the WD progenitor would have started with $\Mzams \approx 6-8~\Msun$. The WD is too massive to have a pure He core so the binary likely experienced 2 phases of CE evolution when the massive companion began H-shell and He-shell burning. This system joins a growing list of massive WDs in close binaries with relatively low-mass ($\Mstar \lesssim 1~\Msun$) companions. Instead of relying on slow AGB winds to remove the stellar envelope, these systems likely ejected several \Msun of material in at least one, possibly two, CE phases. Such systems are extremely useful for placing physical constraints on the complicated processes governing CE evolution and outcomes. Yet \name exemplifies the difficulty in distinguishing between high-mass WDs and low-mass NSs in close binaries, even for bright nearby systems.

\vspace{5mm}
Facilities: UH2.2m (SNIFS); LBT (PEPSI)

Software: astropy \citep{astropy}; numpy \citep{numpy1, numpy2}; matplotlib \citep{matplotlib}; lmfit \citep{lmfit}; scipy \citep{scipy}; spectres \citep{spectres}; emcee \citep{emcee}; pandas \citep{pandas}

\section*{Acknowledgments}

We thank Jennifer Johnson, Marc Pinsonneault, Chris Kochanek, Kris Stanek, Dan Huber, Ben Shappee, and Todd Thompson for useful discussions. 

The LBT is an international collaboration among institutions in the United States, Italy, and Germany. LBT Corporation partners are: The University of Arizona on behalf of the Arizona Board of Regents; Istituto Nazionale di Astrofisica, Italy; LBT Beteili- gungsgesellschaft, Germany, representing the Max-Planck Society, The Leibniz Institute for Astrophysics Potsdam, and Heidelberg University; The Ohio State University, representing OSU, University of Notre Dame, University of Minnesota, and University of Virginia. PEPSI was made possible by funding through the State of Brandenburg (MWFK) and the German Federal Ministry of Education and Research (BMBF) through their Verbundforschung grants 05AL2BA1/3 and 05A08BAC. 

Observations have benefited from the use of ALTA Center (\url{alta.arcetri.inaf.it}) forecasts performed with the Astro-Meso-Nh model. Initialization data of the ALTA automatic forecast system come from the General Circulation Model (HRES) of the European Centre for Medium Range Weather Forecasts.

%




\appendix

\section{Additional Abundance Information}\label{app:obs}

\begin{figure*}
    \centering
    \includegraphics[width=0.95\textwidth]{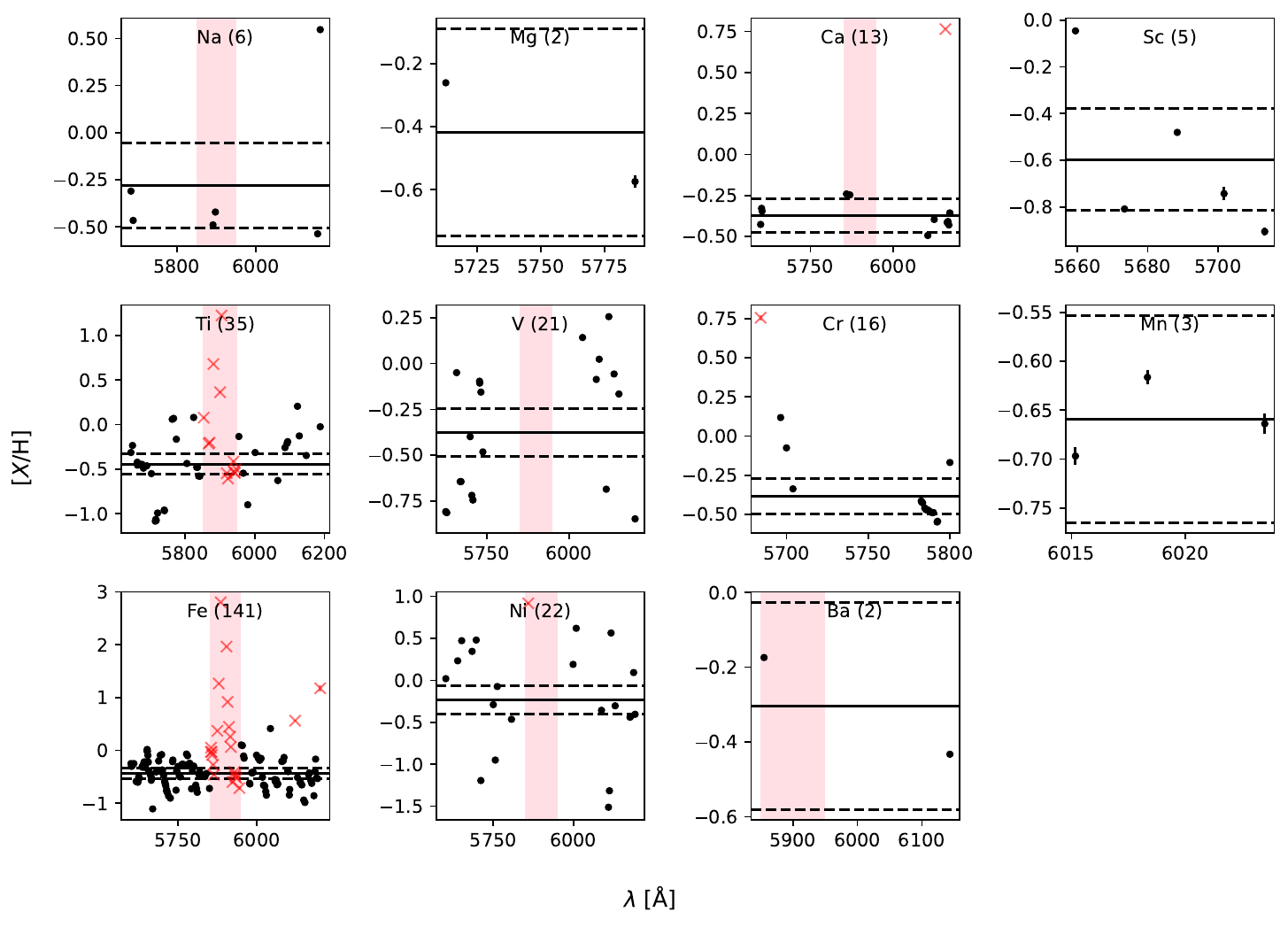}
    \caption{Individual abundance measurements for each element. Solid and dashed black lines represent the mean abundance and its uncertainty (statistical and systematic) which are reported in Table~\ref{tab:korg_abunds}. Those excluded from abundance estimation are marked with red `x's (see \S\ref{subsec:kd.abun}). A pink vertical band marks the region of the strong sodium D absorption lines. Ba and Mg are excluded from our analysis because they only have 2 detected lines.}
    \label{fig:detailed_abunds}
\end{figure*}

Here we provide additional information about the abundance measurements in Table~\ref{tab:korg_abunds}. Fig.~\ref{fig:detailed_abunds} shows the individual abundance measurements for each element, and Table \ref{tab:linelist_sources} contains the associated linelist references.

\begin{table*}[]
    \centering
    \caption{VALD linelist references for the elements measured in \S\ref{subsec:kd.abun}.}
    \label{tab:linelist_sources}

    \begin{tabular}{ll}
    \hline\hline
    Element & References \\ \hline
    Na & \citet{BPM}, \citet{NIST10}, \citet{KP}, \citet{WSG} \\
    Mg & \citet{KP}, \citet{BPM}, \citet{NIST10}, \citet{LZ}\\
    Ca & \citet{K07}, \citet{SR}, \citet{S}, \citet{BPM}, \citet{SN} \\
    Sc & \citet{K09}, \citet{BPM}, \citet{LHSNWC}, \citet{NIST22}, \citet{LD} \\
    Ti & \citet{BPM}, \citet{LGWSC}, \citet{K16}, \citet{Sal12a}, \citet{WLSC}, \\
    &\citet{KL}, \citet{NWL}, \citet{Fors} \\
    V  & \citet{K09}, \citet{LWDFSC}, \citet{TPSa}, \citet{BPM} ,\citet{MFW} \\
    Cr & \citet{K16}, \citet{BPM}, \citet{SLS}, \citet{NIST22}, \\
    &\citet{BA-J}, \citet{LSNE}, \citet{MFW} \\
    Mn & \citet{K07}, \citet{BPM}, \citet{MFW}, \citet{BXPNL}, \\
    &\citet{DLSSC}, \citet{SugCor} \\
    Fe & \citet{K14}, \citet{BPM}, \citet{K13}, \citet{BA-J} \\
    Ni & \citet{K08,WLSCow,LBT,BPM,FMW} \\
    Ba & \citet{CB,MW,BPM} \\
    \hline\hline
    \end{tabular}
\end{table*}

\bibliography{ref, linelist}{}
\bibliographystyle{aasjournal}



\end{document}